Target Capture and Massively Parallel Sequencing of Ultraconserved Elements (UCEs) for

Comparative Studies at Shallow Evolutionary Time Scales

Brian Tilston Smith[1*], Michael G. Harvey[1,2*], Brant C. Faircloth[3], Travis C. Glenn[4], Robb T. Brumfield[1,2]

[1] *Museum of Natural Science, Louisiana State University, Baton Rouge, LA 70803*

[2]*Department of Biological Sciences, Louisiana State University, Baton Rouge, LA 70803*

[3]*Department of Ecology and Evolutionary Biology, University of California, Los Angeles, CA 90095*

[4]*Department of Environmental Health Science, University of Georgia, Athens, GA 30602*

* Authors equally contributed to the manuscript

Corresponding author:

    Brian Tilston Smith

    Telephone: (225) 578-2855

    Email: briantilstonsmith@gmail.com

**ABSTRACT**

Comparative genetic studies of non-model organisms are transforming rapidly due to major advances in sequencing technology. A limiting factor in these studies has been the identification and screening of orthologous loci across an evolutionarily distant set of taxa. Here, we evaluate the efficacy of genomic markers targeting ultraconserved DNA elements (UCEs) for analyses at shallow evolutionary timescales. Using sequence capture and massively parallel sequencing to generate UCE data for five co-distributed Neotropical rainforest bird species, we recovered 776 – 1,516 UCE loci across the five species. Across species, 53 – 77 % of the loci were polymorphic, containing between 2.0 and 3.2 variable sites per polymorphic locus, on average. We performed species tree construction, coalescent modeling, and species delimitation, and we found that the five co-distributed species exhibited discordant phylogeographic histories. We also found that species trees and divergence times estimated from UCEs were similar to the parameters obtained from mtDNA. The species that inhabit the understory had older divergence times across barriers, a higher number of cryptic species within them, and larger effective population sizes relative to the species inhabiting the canopy. Because orthologous UCEs can be obtained from a wide array of taxa, are polymorphic at shallow evolutionary timescales, and can be generated rapidly at low cost, they are an effective genetic marker for studies investigating evolutionary patterns and processes at shallow timescales.

Highly multilocus datasets are becoming more widespread in phylogeography and population genetics. Population genetic theory and simulation studies suggest that increases in the number of unlinked loci included in analyses improve estimates of population genetic parameters (Kuhner et al. 1998; Beerli and Felsenstein 1999; Carling and Brumfield 2007). Under a Wright-Fisher model of evolution, variance in the topology and branch lengths among gene trees is large (Hudson 1990). Processes such as gene flow and natural selection add to this variance. To increase the precision of parameters estimated under the coalescent process, comparative genetic studies have transitioned from single-locus mitochondrial to multilocus data sets (Brumfield et al. 2003). Multilocus data permit the application of parameter-rich coalescent models that can simultaneously estimate multiple demographic parameters (Edwards and Beerli 2000), yet previous methods for collecting multilocus data from non-model taxa were challenging.

The advent of massively parallel sequencing (MPS) and the development of generalized data collection and analytical methods allow biologists to scale data collection to thousands of unlinked loci, even without prior knowledge of the underlying genome sequence (McCormack et al. 2012a). Various approaches use reduced representation libraries for MPS, including amplicon generation (e.g. Morin et al. 2010, O'Neill et al. 2013), restriction-digest based methods (e.g. Baird et al. 2008), transcriptomics (e.g. Geraldes et al. 2011), and sequence capture (e.g. Briggs et al. 2009). Among these, restriction-digest based library preparation methods, such as ddRADs (Peterson et al. 2012), are relatively easy and cheap to perform, and analyses of SNP datasets derived from these methods are increasing (e.g. Emerson et al. 2010, Gompert et al. 2010, McCormack et al. 2012b, Zellmer et al. 2012). However, as the genetic distance among study taxa increases, so does the challenge of collecting data from orthologous loci across species using PCR- or restriction digest-based approaches (Rubin et al. 2012).

The importance of using orthologous loci for comparative phylogeographic studies is unclear. Theory and simulations suggest that population genetic inferences from a non-intersecting set of loci will converge on the same answer, provided the data matrices comprise a sufficient number of independently segregating loci (Kuhner et al. 1998, Beerli and Felsenstein 1999, Carling and Brumfield 2007). It remains an open question, however, whether this is true in practice, especially because mutation rates are known to vary dramatically across the genome (Hodgkinson and Eyre-Walker 2011). As a result, direct comparisons among divergent species are best accomplished by examining sets of orthologous loci.

A recently developed class of markers surrounding ultraconserved DNA elements (UCEs; Faircloth et al. 2012) can be used in conjunction with sequence capture and MPS to generate large amounts of orthologous sequence data among a taxonomically diverse set of species. UCEs are numerous in a diversity of metazoan taxa (Ryu et al. 2012), and over 5000 have been identified in amniotes (Stephen et al. 2008, Faircloth et al. 2012). UCEs have been used to resolve challenging phylogenetic relationships among basal avian (McCormack et al. 2013) and mammalian (McCormack et al. 2012c) groups, to determine the phylogenetic placement of turtles among reptiles (Crawford et al. 2012), and to understand the relationships among ray-finned fishes (Faircloth et al. 2013). Although UCEs are highly conserved across distantly related taxa, their flanking regions harbor variation that increases with distance from the conserved core (Faircloth et al. 2012). The conserved region allows easy alignment across widely divergent taxa, while variation in the flanks is useful for comparative analyses. Faircloth et al. (2012) suggested that because variation within human UCE flanking regions is abundant, UCEs would likely be useful for investigations at shallow evolutionary timescales. Here, we use empirical data from Neotropical birds to demonstrate that UCEs are useful genetic markers at shallow evolutionary timescales.

The lowland Neotropics harbor arguably the highest avian diversity on Earth and the origins of this diversity are contentious (Hoorn et al. 2010; Rull 2011; Brumfield 2012; Ribas et al. 2012; Smith et al. 2012a). The majority of phylogeographic studies on Neotropical birds have used mitochondrial DNA (mtDNA) to estimate divergence times and, in some cases, migration rates. Recent work suggests that ecology may play an important role in avian diversification in the lowland Neotropics (Burney and Brumfield 2009). Highly multilocus datasets in conjunction with highly parameterized models promise to provide further resolution to the processes underlying Neotropical diversity.

Here, we use UCE data to study genetic structure in five widespread Neotropical birds - *Cymbilaimus lineatus, Xenops minutus, Schiffornis turdina, Querula purpurata,* and *Microcerculus marginatus. Cymbilaimus lineatus and Q. purpurata* inhabit the canopy, whereas the other three species inhabit the under- and mid-story. Recent taxonomies treat the understory species as more polytypic than the canopy species. Dickinson (2003), for example, includes 3 subspecies within *C. lineatus* and treats *Q. purpurata* as monotypic, but lists 10, 5, and 6 subspecies for *X. minutus*, *S. turdina*, and *M. marginatus*, respectively. We use target enrichment followed by Illumina sequencing to collect population-level UCE data. In each species, we use these data to (i) infer the species tree, (ii) use demographic models to estimate population genetic parameters, and (iii) identify cryptic species-level diversity within each of the five lineages. We compare and contrast these results with mtDNA (cytochrome *b*) data from the same populations. We examine whether canopy species and understory species display different divergence histories, effective population sizes, and migration rates.

**METHODS**

*Sequence Capture of UCEs from Genetic Samples*

We sampled 40 individuals from five species of widespread Neotropical lowland forest passerine birds - *Cymbilaimus lineatus, Xenops minutus, Schiffornis turdina, Querula purpurata,* and *Microcerculus marginatus* (Table S1; Fig. S1). Understory birds have been shown to exhibit greater genetic differentiation across biogeographic barriers than more vagile canopy birds (Burney and Brumfield 2009). Our taxon sampling allowed us to address how dispersal affects genetic structure by including two canopy (*C. lineatus, Q. purpurata*) and three understory (*X. minutus, S. turdina,* and *M. marginatus*) species. Because each of the five species is distributed across three prominent biogeographic barriers - the Andes Mountains, the Isthmus of Panama, and the upper Amazon River - we were able to examine how these landscape features affected the geographic structuring of populations. For each species, we sampled two individuals from each of four areas of endemism that, collectively, span these barriers: Central America (CA), Chocó (CH), Napo (NA), and Inambari/Rondônia (SA) (Table S1; Fig. 1).

We extracted total genomic DNA from vouchered tissue samples using a DNeasy tissue kit (Qiagen). We fragmented genomic DNA using a BioRuptor NGS (Diagenode) and prepared Illumina libraries using KAPA library preparation kits (Kapa Biosystems) and custom sequence tags unique to each sample (Faircloth and Glenn 2012). To enrich targeted UCE loci, we followed an established workflow (Gnirke et al. 2009; Blumenstiel et al. 2010) incorporating several modifications to the protocol detailed in Faircloth et al. (2012). First, we prepared and sequence-tagged libraries using standard library preparation techniques. Second, we pooled eight samples within each species at equimolar ratios, prior to enrichment, and we blocked the Illumina TruSeq adapter sequence using custom blocking oligos (Table S2). We enriched each pool using a set of 2,560 custom-designed probes (MYcroarray, Inc.) targeting 2,386 UCE loci (see Faircloth et al. 2012 and http://ultraconserved.org for details on probe design). Prior to sequencing, we qPCR-quantified enriched pools, combined pools at equimolar ratios, and sequenced the combined libraries using a partial (90%) lane of a 100-bp paired-end Illumina HiSeq 2000 run (Cofactor Genomics).

We converted BCL files and demultiplexed raw reads using Casava (Illumina, Inc.), and we quality filtered demultiplexed reads using custom Python scripts implemented in the Illumiprocessor workflow (https://github.com/faircloth-lab/illumiprocessor/). This workflow trimmed adapter contamination from reads using SCYTHE (https://github.com/vsbuffalo/scythe), quality-trimmed reads using SICKLE (https://github.com/najoshi/sickle), and excluded reads containing ambiguous (N) bases. Following quality filtering, we generated consensus contigs, *de novo*, for all reads generated for each species using VelvetOptimiser (http://bioinformatics.net.au/software.velvetoptimiser.shtml) and VELVET (Zerbino and Birney 2008). We aligned consensus contigs from VELVET to UCE probes using LASTZ (Harris 2007), and we removed any contigs that did not match probes or that matched multiple probes designed from different UCE loci. We used BWA (Li and Durbin 2009) to generate an index of consensus UCE contigs for each species and we mapped all reads to the index on an individual-by-individual basis. After mapping, we called SNPs and indels for each individual and exported BAM pileups using SAMtools (Li et al. 2009). We also used SAMtools to generate individual-specific consensus sequences for each UCE locus, and we hard-masked low-quality bases (< Q20) within resulting consensus sequences using seqtk (https://github.com/lh3/seqtk) and converted the masked FASTQ files to FASTA format.

We aligned hard-masked FASTA sequences within loci for each species using MAFFT (Katoh et al. 2005), and we used the alignments to generate additional input files for population genetic analyses in lociNGS (Hird 2012). We removed any alignments in a given species that did not contain data for at least one individual from each of the four areas of endemism. For each species, we generated two data matrices – one containing alignments of all loci recovered for the respective species ('single species full' dataset) and another, smaller dataset containing only the subset of loci that we also recovered in the other four species ('single species reduced' dataset). We use these two datasets from each species in most subsequent analyses. To assess variation across species, we also generated interspecific alignments of individuals across all five species in MAFFT - one 'all species full' dataset containing all alignments, and one 'all species reduced' dataset containing only loci with data from at least one individual in each area of endemism in all five species. For clarity, we present details of these datasets in Table 1.

*Summary Statistics*

We used alignments for each species to calculate a range of population genetic summary statistics (haplotype diversity, Watterson's $\theta$, $\pi$, Tajima's *D*) using the program COMPUTE (Thornton 2003). We calculated the number of variable positions, the distance of variable positions from the center of the UCE locus, and the frequency of variation by distance using the "get_smilogram_from_alignments.py" program within the PHYLUCE package (https://github.com/faircloth-lab/phyluce). This program screens locus-specific UCE alignments for variability after excluding masked/ambiguous bases and unknown characters and stores locus- and position-specific variability data in a relational database (http://www.sqlite.org/). The program also tallies the count of variable and total positions by distance from the UCE center to compute a frequency of variable bases by distance and exports these summary data as a CSV file. We plotted these data in R using ggplot2 (Wikham 2009) after masking outlying positions with a frequency > 0.04.

*UCE Species Tree Construction*

We used *BEAST (Heled and Drummond 2010) in the BEAST package (Drummond et al. 2012) to generate species trees using the single species reduced datasets. To assign samples to species/populations in *BEAST, we used the endemic areas Central America (CA), Chocó (CH), Napo (NA), and Inambari/Rondônia (SA) as proxies for populations (Fig. 1). We split the two SA samples of *S. turdina* (Inambari=SA1 and Rondônia=SA2) for this and subsequent analyses because the SA2 sample was east of the Madeira River and in the Rondônia endemic area and showed high divergence from SA1 based on review of the alignments (Fig. S1). We estimated the best-fit model of sequence evolution using CloudForest (https://github.com/ngcrawford/cloudforest). To reduce the number of parameters in our species tree estimation, we used either the JC69 substitution model, by selecting the GTR model with all base frequencies equal and unselecting the ac, ag, cg, and gt operators, or the HKY model with base frequencies set to empirical. We used the following settings for all *BEAST analyses: a strict molecular clock with a fixed rate of 1.0 for all loci, a Yule process on the species tree prior, a lognormal distribution with ($\mu$= 0.001; SD=2.0) for the species.popMean prior and an exponential distribution for the species.yule.birthRate prior ($\mu$=1000). We did not include monomorphic loci in the analyses. We conducted two runs of 2 billion generations for each species and sampled trees every 25,000 generations. We determined the number of burnin replicates to discard and assessed MCMC convergence by examining ESS values and likelihood plots using Tracer v. 1.5 (Rambaut and Drummond 2010). We determined the Maximum Clade Credibility species tree for each species and built cloudograms from the Maximum Clade Credibility gene tree for each locus using DensiTree (Bouckaert 2010).

To compare the species tree topologies from UCEs to an mtDNA gene tree topology, we constructed a single-gene tree of all taxa and estimated divergence times using alignments of cytochrome *b* in the program BEAST v.1.7.4 (Drummond et al. 2012). Details of methods used to generate cytochrome *b* data are available in Supplementary Methods (doi:10.5061/dryad.8fk2s). We used an uncorrelated relaxed molecular clock based on an avian molecular clock (Weir and Schluter 2008) with a lognormal distribution (mean = 0.0105 substitutions/site/million years, SD= 0.1), coalescent constant-size for the tree prior, and a GTR + G substitution model. We ran analyses for 50 million generations and sampled from the posterior distribution every 2500 generations. We validated BEAST analyses by performing multiple independent runs, and we determined the burn-in and assessed MCMC convergence by examining ESS values and likelihood plots in Tracer v.1.5 (Rambaut and Drummond 2010).

*Modeling Demographic History*

We estimated $\tau$, $\theta$, and migration rates using the single species datasets and the Bayesian coalescent program G-PhoCS (Gronau et al. 2011). G-PhoCS is a modified version of the program MCMCcoal (Yang 2002; Rannala and Yang 2003) that allows migration among lineages and integration over all possible haplotypes of unphased diploid genotypes. G-PhoCS requires a user-specified topology to estimate model parameters, thus we used the *BEAST populations and species tree topology for each species. For each species, we ran four models examining the impact of migration and the inclusion of loci not present in all five species on parameter estimates. The models were: (m$_1$) single species reduced, no migration; (m$_2$) single species reduced, migration; (m$_3$) single species full, no migration; and (m$_4$) single species full, migration.

G-PhoCS uses a gamma ($\alpha$,$\beta$) distribution to specify the prior distribution for the population standardized mutation rate parameter ($\theta = 4N_e\mu$ for a diploid locus, where $\mu$ is per nucleotide site per generation), the species divergence time parameter ($\tau = T\mu$; $T$ = species divergence time in millions of years), and the migration rate per generation parameter ($m_{sx} \times \theta_x/4 = M_{sx}$), which is the proportion of individuals in population $x$ that arrived by migration from population $s$ per generation. We evaluated a series of prior distributions ranging from shallow to deep divergences, small to large theta values, and low to high migration rate by changing the shape parameter ($\alpha$) and scale parameter ($\beta$), which have a mean $\alpha/\beta$ and variance $s^2 = \alpha/\beta^2$. For the final analyses, we used two different priors for $\tau$-$\theta$: (1, 30) and (1,300). These priors represent wide ranges of divergence times and effective population sizes. The $\tau$-$\theta$ prior distributions had no apparent influence on the posterior estimates of $\tau$ and extant $\theta$, except in one case where there was uncertainty surrounding the ancestral theta of NA+SA in *C. lineatus*. For the migration rate prior, when we increased the mean of the migration rate prior distribution, posterior estimates of migration rate had larger uncertainty, poorer mixing, and in some cases runs failed. Despite these problems, the medians of the posterior estimates in the runs that finished were qualitatively similar across the different prior distributions we tested. Thus, to achieve adequate convergence we set the migration rate prior $\alpha$ and $\beta$ to 1.0, 10. We ran each analysis twice for 1,000,000 generations and collected samples from the posterior distribution every 500 generations. We assessed MCMC convergence and determined burn-in by examining ESS values and likelihood plots in the program Tracer v.1.5 (Rambaut and Drummond 2010).

In the absence of a fossil or geologic calibration, we converted raw parameter estimates using a relative substitution rate (see MCMCcoal manual) by calculating π, the average pairwise genetic distance within each species, for each UCE locus and the mtDNA gene cytochrome *b*. We averaged π across UCEs and scaled the UCE π / cytochrome *b* π ratio to the cytochrome *b* rate of 0.0105 substitutions/site/million years, which is based on multiple independent geologic and fossil calibrations (Weir and Schluter 2008).

*mtDNA Divergence Times*

To compare UCE divergence time estimates to those estimated from mtDNA, we performed coalescent modeling using only the mtDNA gene cytochrome *b*. We attempted to use G-PhoCS (Gronau et al. 2011) to estimate model parameters, but runs did not converge, perhaps because G-PhoCS is designed for genomic data rather than a single locus. Instead, we used the Bayesian coalescent program MCMCcoal (Rannala and Yang 2003) because the method implemented is similar to that in G-PhoCS (Gronau et al. 2011).

MCMCcoal implements the Jukes Cantor substitution model and assumes no gene flow among species. As with G-PhoCS, MCMCcoal uses a gamma (α,β) distribution to specify the prior distribution for the population standardized mutation rate parameter and the species divergence time parameter. We used an inheritance scalar (0.25) to account for the four-fold difference in effective population size between nuclear DNA and mtDNA. We performed several preliminary runs with various τ prior distributions and the results suggested that the specified prior did not strongly influence the posterior. In contrast the specified θ prior distributions influenced the posterior. For the final analyses, we used the same two prior distributions from our G-PhoCS analyses: (1,30) and (1,300). Once again, we assumed a generation time of one year. For the topology of each species, we used the results from our \*BEAST analyses. We analyzed each species independently and we ran each prior combination twice for 1,000,000 generations, sampled every five generations and set the burn-in to 50,000 generations.

*Bayesian Species Delimitation*

Current species limits in tropical birds often underrepresent actual species diversity because many species' taxonomies have yet to be evaluated with genetic methods (Tobias et al. 2008; Smith et al. 2013). To delimit the number of species within each of the five taxa, we used the program BP&P v 2.0b (Rannala and Yang 2003; Yang and Rannala 2010), which implements a Bayesian modelling approach to generate speciation probabilities of closely related taxa from multilocus sequence data. The program takes into account gene tree uncertainty and lineage sorting and assumes no gene flow among species after divergence, no recombination among loci, and the JC69 finite-sites substitution model. Included in the speciation model are the divergence time parameter ($\tau$) and population standardized mutation rate parameter ($\theta = 4N_e\mu$, where $N_e$ is the effective population size and $\mu$ is the substitution rate per site per generation). To generate BP&P input files, we modified ".ima" files for each species to fit the formatting specifications of the program. BP&P generates a posterior distribution of speciation models containing differing numbers of species and estimates speciation probabilities from the sum of the probabilities of all models for speciation events at each node in the guide tree. To specify the guide tree for each species, we input a fixed species tree (rjMCMC algorithm 0) using the topologies estimated by \*BEAST, and we performed species delimitation using the reduced dataset. Because BP&P has been shown to be sensitive to the choice of prior distributions (Zhang *et al.* 2011), we performed analyses using a range of priors that represented different population sizes and different ages for the root in the species tree (Leaché and Fujita 2010). BP&P uses a gamma distribution where the shape parameter (α) and scale parameter (β) are specified by the user. For the final analyses, we used the same prior distributions as our G-PhoCS and MCMCcoal analyses: (1,30) and (1,300) and assessed convergence using the same diagnostics as our previous analyses. We ran analyses with each prior twice for 1,000,000 generations, sampling every five, and specified a burnin of the first 50,000 generations.

**RESULTS**

The partial Illumina HiSeq lane produced 255 million reads from the 40 samples following initial processing with Illumiprocessor. Four of the eight samples of *M. marginatus* (LSUMZ 2460, 9127, 11839, 106784) failed and we excluded these from additional analyses. From the remaining 36 samples we input 249 million trimmed reads into the assembly process (Tables S1, S3), and assembled a total of 29 million reads into contigs across the five study species (Table S3). Across species, mean contig length varied from 258 to 371 bp, and mean read depth varied from 182× to 387× (Table S3). Between 951 and 1,796 contigs aligned to UCE loci in each species, and we retained 812 to 1,580 of these in the MAFFT alignments (Table S3). Many alignments did not contain all individuals – each individual was present in between 560 and 1,499 alignments (Table S1).

After filtering out poor alignments with limited overlapping sequence and alignments containing less than one individual in each population, 776 to 1,516 loci remained in the single species full datasets for the five study species (Table 1; Table S3). From these, we recovered 166 loci shared across all five species, and these comprised the single species reduced datasets. For the interspecific alignments, we recovered 2,219 total UCE loci (all species full dataset), of which 169 contained at least one individual in each area of endemism across all species (all species reduced dataset; Table 1).

*UCE Summary Statistics*

The percentage of polymorphic loci in each species ranged from 53 to 77%, and the average number of variable sites in each polymorphic locus ranged from 2.0 to 3.2 (Table S4). Values of Watterson's θ and π averaged across UCEs were lowest in *Q. purpurata* and *M. marginatus*, and highest in *S. turdina* (Table 2; Fig. S2). Values of Tajima's *D* within species ranged from −0.22 ± 0.86 to −0.63 ± 0.63, although there was considerable variation across loci (Table 2; Fig. S2).

Mean values of Watterson's θ, π, and Tajima's *D* differed between the single species full and single species reduced datasets in only 3 of 15 independent comparisons (Mann-Whitney *U*-Test *P* < 0.05; Table 2; Fig. S2). Summary statistics calculated from cytochrome *b* were outside the confidence limits of the equivalent statistics calculated from UCEs (single species full datasets) in 12 of 15 independent comparisons (Table 2). Variation in summary statistic estimates decreased as the number of UCE loci used increased (Fig. S3). Plots of variation versus alignment position for each species indicated that the frequency of variant bases increased with distance from the center of UCEs (Fig. 2; Fig S4). Plots containing all taxa showed that the frequency of variant bases reached an asymptote around 300 bp from the UCE centers and subsequently decreased. This decrease may occur because fewer alignments extend past 300 bp from center and because those alignments having data in outer flanks contain fewer individuals (Fig. S5). The decrease in variability in outer flanks may also result from reduced read depth across these outer positions.

*Species Trees*

The best-fit substitution model for the majority of UCEs across species was JC69 (Fig. S6). Species tree estimation using the single species reduced datasets required between 17 and 27 days (Intel Core i7 3.4 GHz) and the runs converged and ESS values were > 200. Species tree estimation recovered generally highly supported topologies (Fig. 3a; Fig. S7) and discordance across individual gene tree topologies (Fig. 3b). The only parameters with ESS values < 200 were four population size parameters in *C. lineatus*, which we attribute to an unresolved node in that tree. Species trees were highly congruent in topology with mtDNA gene trees from the same species (Fig. S8). Species trees for *X. minutus*, *Q. purpurata,* and *M. marginatus* were fully resolved with high support values for all nodes (posterior probabilities, PP = 1.0). In these three species there was a basal divergence across the Andes separating a clade composed of Central America (CA) and Chocó (CH) populations west of the Andes from a clade comprising all other lineages found east of the Andes. In *X. minutus* and *Q. purpurata*, the Napo (NA) and Inambari/Rondônia (SA) populations were sister. The topology for *C. lineatus* was similar, with a basal divergence that represented a cross-Andes break. Although the trans-Andean clade (CA+CH) clade (PP = 1.0) was well-supported in C. lineatus, the cis-Andean clade (SA+NA) was less well supported (PP = 0. 70) with some support for an alternative topology with NA sister to the trans-Andean clade (Fig. S7). The species tree for *S. turdina* was fully resolved with high support values (PP > 0.99; Fig. 5), with the two SA populations (SA1 = Inambari and SA2 = Rondônia) sister (PP = 1.0) and NA sister to the (CA+CH) clade (PP = 1.0).

*Comparative Divergence Estimates*

G-PhoCS run lengths varied from 100 to 150 hours (Intel Core i7 3.4 GHz) and all runs converged, produced similar results using different priors, and ESS values for all parameters were >1000. The inclusion of migration and the analysis of full versus reduced UCE datasets across species had minimal effects on parameter estimates (Fig. 4; Figs. S9 − S10; Table S7). Parameter estimates were similar across all four models ($m_1 − m_4$) examined in G-PhoCS. Tau ($\tau$) estimates (from model $m_4$) across the Andes suggest different divergences across the five species: *C. lineatus* with means (95% HPD intervals) = $2.64 \times 10^{-4}$ ($2.40 − 3.00 \times 10^{-4}$); *X. minutus* = $1.04 \times 10^{-3}$ ($9.80 \times 10^{-4} − 1.08 \times 10^{-3}$); *S. turdina* = $5.63 \times 10^{-4}$ ($5.00 − 6.10 \times 10^{-4}$); *Q. purpata* = $4.19 \times 10^{-4}$ ($4.00 − 4.50 \times 10^{-4}$); and *M. marginatus* = $7.42 \times 10^{-4}$ ($6.80 − 8.10 \times 10^{-4}$). After converting $\tau$ parameters to time in millions of years using relative substitution rates, we found that understory species had older divergences across the Andes than canopy species (Fig. 4a). Divergences times across the Isthmus of Panama and Amazon River were also older in understory species than canopy species (Fig 4a). In comparison to the UCE estimates of divergence times, eight of the mtDNA divergence time estimates were older, three were younger, and four were approximately the same (Fig. 4a). For the majority of divergences, the mtDNA time estimates had larger credible intervals than the estimates from the UCE datasets (Fig. 4a).

*Effective Population Sizes and Migration*

Theta and effective population sizes estimated from UCEs provided insight into demographic patterns between understory and canopy birds (Fig. 4b; Figs. S9 − S10; Tables S6, S8). We converted $\theta$ values ($\theta = 4N_e\mu$) using relative substitution rates (Table S5) and a generation time of one year, and we found that understory species (*X. minutus*, *S. turdina*, and *M. marginatus*) had larger effective population sizes than canopy species (*C. lineatus* and *Q.*

*purpurata*) in all models (Fig. 4b ; Fig. S9). $N_e$ estimates were similar with and without the inclusion of migration in the model (Fig. S9), and $N_e$ estimated from the single species reduced datasets produced values similar to those estimated from the single species full datasets (Fig. 4b).

Migration rates were non-zero in all analyses including migration, but migration rate estimates were sensitive to the prior distribution used. Understory species had higher mean values of migration parameters than canopy species, but credible intervals were large and included values near zero for all migration estimates (Fig. S10). Inferred migration estimates were similar using both the single species reduced and single species full datasets (Fig. S10).

*Species Delimitation*

BP&P run lengths varied from 120 to 200 hours (Intel Core i7 3.4 GHz) with the independent runs producing similar results and all runs converging with high ESS values. The analyses were stable among independent runs and the different priors produced similar results for most speciation events. The number of inferred species within each species from the BP&P analysis varied from two to four: *C. lineatus* = 2, *X. minutus* = 4, *S. turdina* = 5, *Q. purpurata* = 2, and *M. marginatus* = 2. Speciation probabilities were sensitive to prior distributions for some nodes in *S. turdina* and *C. lineatus* (Fig S11). Overall, speciation probabilities were higher in the understory species than in the canopy species (Fig. S11).

**DISCUSSION**

Using target enrichment and MPS of UCE loci distributed across the genome, we generated alignments of 166 loci shared among five co-distributed bird species ('reduced' datasets). We found that most loci were polymorphic within species, and we were able to use the enriched sequence data to infer species trees, perform species delimitation analyses, and reconstruct demographic histories for all species. We also generated larger (776 to 1,516 loci) 'full' datasets that contained both the shared loci and loci that we recovered from four of fewer species. We used these datasets to reconstruct demographic histories for all species. The UCE results presented here complement previous studies that highlighted the utility of UCEs as a source of informative phylogenetic markers (Crawford et al. 2012; Faircloth et al. 2012; Faircloth et al. 2013, McCormack et al. 2012c; McCormack et al. 2013), and extend the application of UCEs to studies of comparative phylogeography at the species and population level. The success of UCEs in both population genetic inference and in resolving phylogenetic relationships at deep and shallow evolutionary time scales confirms their utility in both micro- and macro-evolutionary investigations.

*Data Generation, UCE Variability, and Computation*

We assembled 11.8% of the trimmed reads into contigs and only 27.4% of assembled contigs mapped to UCEs, suggesting that we recovered a high proportion of off-target reads. We were able to assemble nearly complete mitochondrial genomes for most samples using raw read data, suggesting that many of the off-target reads represent mtDNA "contamination". This contamination may be due to the high copy number of mitochondria in the tissues we used. It is remarkable that despite this limitation, we were able to obtain data matrices of 166 shared loci from all five species. Looking forward, improved enrichment efficiency due to optimized lab protocols should lead to better capture of UCEs, and technological improvements in MPS may result in longer reads producing longer contigs.

Levels of variation in the flanking regions of UCEs are comparable to those found in sequences and introns recovered from Sanger sequencing of nuclear DNA within species. We found that 53-77% of UCEs in each species were polymorphic and UCEs contained $2.0 - 3.2$ SNPs on average within species, or an average of one SNP every $222 - 550$ bp. These SNP discovery rates are similar to intraspecific studies that found, for example, one SNP per 142 (Geraldes et al. 2011) or 518 (Bruneaux et al. 2012) bp, and 1 polymorphism every $16 - 157$ bp in introns (Smith and Klicka 2013). Although $2.0 - 3.2$ SNPs per locus may result in low gene tree resolution, methods that integrate low information content across many loci, along with longer contigs due to MPS improvements, may further improve the utility of UCEs for future studies.

To examine the effects of the number of loci on parameter estimation, we compared summary statistics estimated from subsamples of the UCE datasets containing different numbers of loci. As expected, confidence intervals generally decreased as the number of UCE loci used in the analysis increased (Fig. S3). We also compared population genetic parameters estimated from the single species reduced datasets (166 loci) to those estimated from the single species full datasets (776 - 1,516 loci) and mtDNA. Again, credible intervals generally decreased as the number of loci used to estimate parameters increased (Fig. 4). The single species full and single species reduced UCE datasets produced similar summary statistics and mean parameter estimates, and the chronologic sequence of divergence events across the five species was consistent between datasets.

It remains unclear how important it is that large UCE data matrices be composed exclusively of the same loci when drawing comparisons across species or datasets. Parameters estimated in G-PhoCS from the single species reduced and single species full datasets were similar. There were analytical limitations that did not allow us to compare parameters estimated from both data sets in *BEAST and BP&P. For some analyses (e.g. G-PhoCS) we were able to efficiently and rapidly analyze the single species full data sets (~100 hours on a 3.4 GHz Intel Core I7 processor). In contrast, preliminary analyses using the single species full data sets in *BEAST and BP&P indicated that runs would take weeks to months to finish. Our results suggest that datasets composed of large numbers of non-intersecting loci may be sufficient for many comparative phylogeographic applications, but additional data and faster computational methods are needed to explore this further.

With highly multilocus UCE data we were able to estimate θ with relatively high precision relative to mtDNA, but credible intervals on migration rates were large, highlighting the general difficulty of measuring dispersal parameters (Broquet and Petit 2009). Population divergence times estimated from mtDNA tended to be older than those estimated from UCEs (Fig. 4), consistent with the idea that, because gene trees are embedded within species trees, analyses based on single gene trees will overestimate divergence times (Edwards and Beerli 2000). Three divergence estimates are shallower in the mtDNA estimate than the multilocus UCE estimate (Fig. 4), implicating either recent mitochondrial gene flow or the fact that, in rare cases, gene divergence is shallower than the confidence intervals associated with population divergence (Edwards and Beerli 2000). Information on theta and migration between populations in highly multilocus datasets, such as those generated using UCEs, provides another layer of information that complements what has been available historically using mtDNA.

*Comparative Phylogeography in Neotropical Forest Birds*

Our phylogeographic analyses based on UCEs indicated that co-distributed taxa exhibit discordant evolutionary histories. Divergence time patterns inferred from UCEs are consistent with recent studies that show multiple speciation events across the Andes (Brumfield and Edwards 2007; Miller et al. 2008), Amazonian rivers (Naka et al. 2012), and the Isthmus of Panama (Smith et al. 2012b). Previous comparative phylogeographic studies on Neotropical birds have not assessed the influence of gene flow on inferred patterns. However, the inclusion of migration in our coalescent modeling had a negligible influence on divergence time estimates for these species. Since our migration rates were sensitive to the prior distribution and substitution rates for UCEs were low, it is unclear whether the consistency in divergence times across our models represents limited gene flow across barriers or insufficient information in the UCE datasets to accurately estimate gene flow. Despite uncertainty around the robustness of the migration rate estimates, the inclusion of migration in the coalescent modeling allowed us to relax the pure isolation assumption. Based on the BP&P analyses, understory species typically had higher speciation probabilities or more inferred species. Some nodes had low (PP < 0.95) speciation probabilities, likely because of gene flow among endemic areas and/or shallow divergences among areas. Overall, divergence time and species delimitation estimates were largely concordant with the results of a previous mtDNA study that showed greater genetic differentiation in poorly dispersing understory species than in more vagile canopy species (Burney and Brumfield 2009).

These UCE data provide the first genomic insight into the effective population sizes of rainforest birds, a critical parameter for inferring how species have evolved in response to past events. We found that understory birds (*X. minutus, S. turdina,* and *M. marginatus*) have larger effective population sizes than species that inhabit the canopy (*C. lineatus* and *Q. purpurata*). This finding, if corroborated with a larger sample of species, has important implications for understanding how species ecology influences genetic differentiation, turnover rates among ecological guilds, and the accumulation of diversity over time. Canopy species have larger territory sizes and occur at lower densities than understory species (Munn and Terborgh 1985; Terborgh et al. 1990). Canopy species may also have more dynamic or unstable distributions and population sizes (Greenberg 1981; Loiselle 1988), particularly *Q. purpurata*, which is largely frugivorous and may track ephemeral sources of fruit across the landscape. The observation that *C. lineatus* and *Q. purpurata* have less divergent populations is consistent with the expectation that species with small or unstable population sizes will be less likely to accumulate genetic diversity over time (Leffler et al. 2012). Our results provide additional preliminary evidence that ecology is important in predicting the population histories of tropical birds. As precision in the estimation of population genetic parameters improves with multilocus data, so too will our ability to elucidate the linkage between ecological and evolutionary processes.

**Acknowledgements**

We thank D. Dittmann (LSUMNS), J. Klicka (UWBM), M. Robbins (KU), and A. Aleixo (MPEG) for providing samples, S. Herke for assisting with laboratory work, L. Yan for assistance with *de novo* assembly on the LSU HPC Philip cluster, and I. Gronau for help running G-PhoCS. We thank S. Hird for writing custom scripts to process our data in lociNGS. We thank A. Cuervo and E. Rittmeyer for providing helpful assistance during data generation and analysis. F. Anderson, M. Charleston, and two anonymous reviewers provided comments that greatly improved this manuscript.  This research was supported by NSF grant DEB-0841729 to RTB.

AVAILABILITY

Supplementary material, including data files and/or online-only appendices, can be found in the Dryad data repository at http://datadryad.org, doi:10.5061/dryad.8fk2s.

REFERENCES

Baird N.A., Etter P.D., Atwood T.S., Currey M.C., Shiver A.L., Lewis Z.A., Selker E.U., Cresko W.A., Johnson, E.A. 2008. Rapid SNP discovery and genetic mapping using sequenced RAD markers. PLoS One 3:e3376.

Beerli P., Felsenstein J. 1999. Maximum-likelihood estimation of migration rates and effective population numbers in two populations using a coalescent approach. Genetics 152:763-773.

Blumenstiel B., Cibulskis K., Fisher S., DeFelice M., Barry A., Fennell T., Abreu J., Minie B., Costello M., Young G., Maguire J., Melnikov A., Rogov P., Gnirke A., Gabriel S. 2010. Targeted exon sequencing by in-solution hybrid selection. Curr. Protoc. Hum. Genet. Chapter 18: Unit 18.4.

Bouckaert R.R. 2010. DensiTree: making sense of sets of phylogenetic trees. Bioinformatics 26:1372-1373.

Briggs A.W., Good J.M., Green R.E., Krause J., Maricic T., Stenzel U., Lalueza-Fox C., Rudan P., Brajkovi D., Ku an Ž. 2009. Targeted retrieval and analysis of five Neandertal mtDNA genomes. Science 325:318-321.

Broquet T., Petit E. 2009. Molecular estimation of dispersal for ecology and population genetics. Annu. Rev. Ecol. Evol. Syst. 40:193-216.

Brumfield R.T. 2012. Inferring the origins of lowland Neotropical birds. Auk 129:367-376.

Brumfield R.T., Beerli P., Nickerson D.A., Edwards S.V. 2003. The utility of single nucleotide polymorphisms in inferences of population history. Trends Ecol. Evol. 18: 249-256.

Brumfield R.T., Edwards S.V. 2007. Evolution into and out of the Andes: a Bayesian analysis of historical diversification in *Thamnophilus* antshrikes. Evolution 61:346-367.

Bruneaux M., Johnston S.E., Herczeg G., Merilä J., Primmer C.R., Vasemägi A. 2013. Molecular evolutionary and population genomic analysis of the nine-spined stickleback using a modified restriction-site-associated DNA tag approach. Mol. Ecol. 22: 565-582.

Burney C.W., Brumfield R.T. 2009. Ecology predicts levels of genetic differentiation in Neotropical birds. Am. Nat. 174:358-368.

Carling M.D., Brumfield R.T. 2007. Gene sampling strategies for multi-locus population estimates of genetic diversity (theta). PLoS One 2:e160.

Crawford N.G., Faircloth B.C., McCormack J.E., Brumfield R.T., Winker K., Glenn T.C. 2012. More than 1000 ultraconserved elements provide evidence that turtles are the sister group of archosaurs. Biol. Lett. 8:783-786.

Dickinson E.C, editor. 2003. The Howard and Moore complete checklist of the birds of the World. 3rd ed. Princeton (NJ): Princeton University Press.

Drummond, A. J., Suchard, M. A., Xie, D., Rambaut, A. 2012. Bayesian phylogenetics with BEAUti and the BEAST 1.7. Mol. Biol. Evol. 29: 1969-1973.

Edwards S.V., Beerli P. 2000. Perspective: Gene divergence, population divergence, and the variance in coalescence time in phylogeographic studies. Evolution 54:1839-1854.

Emerson K.J., Merz C.R., Catchen J.M., Hohenlohe P.A., Cresko W.A., Bradshaw W.E., Holzapfel C.M. 2010. Resolving postglacial phylogeography using high-throughput sequencing. Proc. Natl. Acad. Sci. USA 107:16196-16200.

Faircloth B.C., Glenn T.C. 2012. Not all sequence tags are created equal: designing and
     validating sequence identification tags robust to indels. PLoS One 7:e42543.

Faircloth B.C., McCormack J.E., Crawford N.G., Harvey M.G., Brumfield R.T., Glenn T.C.
     2012. Ultraconserved elements anchor thousands of genetic markers spanning multiple
     evolutionary timescales. Syst. Biol. 61:717-726.

Faircloth B.C., Sorenson L., Santini F., Alfaro M.E. 2013.  A phylogenomic perspective on the
     radiation of ray-finned fishes based upon targeted sequencing of ultraconserved elements
     (UCEs).  PLoS One. DOI:10.1371/journal.pone.0065923

Geraldes A., Pang J., Thiessen N., Cezard T., Moore R., Zhao Y., Tam A., Wang S., Friedmann
     M., Birol I. 2011. SNP discovery in black cottonwood (*Populus trichocarpa*) by population
     transcriptome resequencing. Mol. Ecol. Res. 11:81-92.

Gnirke A., Melnikov A., Maguire J., Rogov P., LeProust E.M., Brockman W., Fennell T.,
     Giannoukos G., Fisher S., Russ C., Gabriel S., Jaffe D.B., Lander E.S., Nusbaum C. 2009.
     Solution hybrid selection with ultra-long oligonucleotides for massively parallel targeted
     sequencing. Nat. Biotechnol. 27:182-189.

Gompert Z., Forister M.L., Fordyce J.A., Nice C.C., Williamson R.J., Buerkle C.A. 2010.
     Bayesian analysis of molecular variance in pyrosequences quantifies population genetic
     structure across the genome of *Lycaeides* butterflies. Mol. Ecol. 19:1473-2455.

Greenberg R. 1981. The abundance and seasonality of forest canopy birds on Barro-Colorado
     Island, Panama. Biotropica 13:241–251.

Gronau I., Hubisz M.J., Gulko B., Danko C.G., Siepel A. 2011. Bayesian inference of ancient
     human demography from individual genome sequences. Nat. Genet. 43:1031-1034.

Harris R.S. 2007. Improved pairwise alignment of genomic DNA. State College (PA): The Pennsylvania State University.

Heled J., Drummond A.J. 2010. Bayesian inference of species trees from multilocus data. Mol. Biol. Evol. 27:570-580.

Hird S.M. 2012. lociNGS: a lightweight alternative for assessing suitability of next-generation loci for evolutionary analysis. PLoS One 7:e46847.

Hodgkinson A., Eyre-Walker A. 2011. Variation in the mutation rate across mammalian genomes. Nat. Rev. Genet. 12:756-766.

Hoorn C., Wesselingh F.P., ter Steege H., Bermudez M.A., Mora A., Sevink J., Sanmartín I., Sanchez-Meseguer A., Anderson C.L., Figueiredo J.P., Jaramillo C., Riff D., Negri F.R., Hooghiemstra H., Lundberg J., Stadler T., Särkinen T., Antonelli A. 2010. Amazonia through time: Andean uplift, climate change, landscape evolution, and biodiversity. Science 330:927-931.

Hudson R. 1990. Gene genealogies and the coalescent process. In: Futuyma D., Antonovics J., editors. Oxford surveys in evolutionary biology, vol. 7. New York: Oxford University Press. p. 1-44.

Katoh, K., Kuma K.-i., Toh H., Miyata T. 2005. MAFFT version 5: improvement in accuracy of multiple sequence alignment. Nucleic Acids Res. 33:511-518.

Kuhner M. K., Yamato J., Felsenstein J. 1998. Maximum likelihood estimation of population growth rates based on the coalescent. Genetics 149:429-434.

Leaché A.D., M. K. Fujita. 2010. Bayesian species delimitation in West African forest geckos (*Hemidactylus fasciatus*). Proc. R Soc. B 277:3071-3077.

Leffler E.M., Bullaughey K., Matute D.R., Meyer W.K., Ségurel L., Venkat A., Andolfatto P., Przeworski M. 2012. Revisiting an old riddle: What determines genetic diversity levels within species? PLoS Biol. 10:e1001388.

Li H., Durbin R. 2009. Fast and accurate short read alignment with Burrows-Wheeler transform. Bioinformatics 25:1754-1760.

Li H., Handsaker B., Wysoker A., Fennell T., Ruan J., Homer N., Marth G., Abecasis G., Durbin R. 2009. The Sequence Alignment/Map format and SAMtools. Bioinformatics 25:2078-2079.

Loiselle B.A. 1988. Bird abundance and seasonality in a Costa Rican lowland forest canopy. Condor 90:761–772.

McCormack J.E., Faircloth B.C., Crawford N.G., Gowaty P.A., Brumfield R.T., Glenn T.C. 2012c. Ultraconserved elements are novel phylogenomic markers that resolve placental mammal phylogeny when combined with species-tree analysis. Genome Res. 22:746-754.

McCormack J.E., Harvey M.G., Faircloth B.C., Crawford N.G., Glenn T.C., Brumfield R.T. 2013. A phylogeny of birds based on over 1,500 loci collected by target enrichment and high-throughput sequencing. PLoS One.

McCormack J.E., Hird S.M., Zellmer A.J., Carstens B.C., Brumfield R.T. 2012a. Applications of next-generation sequencing to phylogeography and phylogenetics. Mol. Phylogenet. Evol. doi:10.1016/j.ympev.2011.12.007

McCormack J.E., Maley J.M., Hird S.M., Derryberry E.P., Graves G.R., Brumfield R.T. 2012b. Next-generation sequencing reveals phylogeographic structure and a species tree for recent bird divergences. Mol. Phylogenet. Evol. 62:397-406.

Miller M.J., Bermingham E., Klicka J., Escalante P., Raposo do Amaral F.S., Weir J.T., Winker K. 2008. Out of Amazonia again and again: episodic crossing of the Andes promotes diversification in a lowland forest flycatcher. Proc. R Soc. B 275:1133-1142.

Morin P.A., Archer F.I., Foote A.D., Vilstrup J., Allen E.E., Wade P., Durban J., Parsons K., Pitman R., Li L.. 2010. Complete mitochondrial genome phylogeographic analysis of killer whales (*Orcinus orca*) indicates multiple species. Genome Res. 20:908-916.

Munn C. A., Terborgh J. 1985. Permanent canopy and understory flocks in Amazonia: species composition and population density. In Buckley P.A., Foster M.S., Morton E.S., Ridgely R.S., Buckley F.G., editors. Neotropical Ornithology. Ornithological Monographs Number 36. Washington (DC): American Ornithologists Union. Pages 683-712

Naka L.N., Bechtoldt C.L., Henriques L.M.P., Brumfield R.T. 2012. The role of physical barriers in the location of avian suture zones in the Guiana Shield, northern Amazonia. Am. Nat. 179:E115-E132.

O'Neill E.M., Schwartz R., Bullock C.T., Williams J.S., Shaffer H.B., Aguilar-Miguel X., Parra-Olea G., Weisrock D.M. 2013. Parallel tagged amplicon sequencing reveals major lineages and phylogenetic structure in the North American tiger salamander (*Ambystoma tigrinum)* species complex. Mol. Ecol. DOI: 10.111/mec.12049

Peterson B. K., Weber J. N., Kay E. H., Fisher H. S., Hoekstra H. E. 2012. Double digest RADseq: An inexpensive method for de novo snp discovery and genotyping in model and non-model species. PLoS One 7:e37135.

Rambaut A., Drummond A.J. 2010. Tracer v 1.5. Program distributed by the authors. Oxford: University of Oxford.

Rannala, B., Yang Z. 2003. Bayes estimation of species divergence times and ancestral population sizes using DNA Sequences from multiple loci. Genetics 164:1645-1656.

Ribas C.C., Aleixo A., Nogueira A.C.R., Miyaki C.Y., Cracraft J. 2012. A palaeobiogeographic model for biotic diversification within Amazonia over the past three million years. Proc. R Soc. B 279: 681-689.

Rubin B.E.R., Ree R.H., Moreau C.S. 2012. Inferring phylogenies from RAD sequence data. PLoS One 7:e33394.

Rull V. 2011. Neotropical biodiversity: timing and potential drivers. Trends in Ecology and Evolution 26:508-513.

Ryu T., Seridi L., Ravasi T. 2012. The evolution of ultraconserved elements with different phylogenetic origins. BMC Evol. Biol. 12:236.

Smith B.T., Amei A., Klicka J. 2012b. Evaluating the role of contracting and expanding rainforest in initiating cycles of speciation across the Isthmus of Panama. Proc. R. Soc. B 279:3520-3526.

Smith B.T., Bryson R.W., Houston D., Klicka J. 2012a. An asymmetry in niche conservatism contributes to the latitudinal species diversity gradient in New World vertebrates. Ecol. Lett. 15:1318-1325.

Smith B.T., Klicka J. 2013. Examining the role of effective population size on mitochondrial and multilocus divergence time discordance in a songbird. PLoS One. 8(2): e55161.

Smith B.T., Ribas C.C., Whitney B.W., Hernández-Baños B.E., Klicka J. 2013. Identifying biases at different spatial and temporal scales of diversification: a case study using the Neotropical parrotlet genus *Forpus*. Mol. Ecol. 22: 483-494.

Stephen S., Pheasant M., Makunin I.V., Mattick J.S. 2008. Large-scale appearance of
    ultraconserved elements in tetrapod genomes and slowdown of the molecular clock. Mol.
    Biol. Evol. 25:402-408.

Terborgh J., Robinson S.K., Parker III T.A., Munn C.A., Pierpont N. 1990. Structure and
    organization of an Amazonian forest bird community. Ecol. Monogr. 60:213-238.

Thornton, K. 2003. libsequence: a C++ class library for evolutionary genetic analysis.
    Bioinformatics 19:2325-2327.

Tobias J.A., Bates J.M., Hackett S.J., Seddon N. 2008. Comment on "The Latitudinal Gradient in
    Recent Speciation and Extinction Rates of Birds and mammals". Science, 901c.

Weir J.T., Schluter D. 2008. Calibrating the avian molecular clock. Mol. Ecol. 17:2321-2328.

Wikham H. 2009. ggplot2: elegant graphics for data analysis. New York: Springer.

Yang Z. 2002. Likelihood and Bayes estimation of ancestral population sizes in hominoids using
    data from multiple loci. Genetics 162:1811-1823.

Yang Z., Rannala B. 2010. Bayesian species delimitation using multilocus sequence data. Proc.
    Natl. Acad. Sci. U.S.A. 107:9264-9269.

Zellmer A.J., Hanes M.M., Hird S.M., Carstens B.C. 2012. Deep phylogeographical structure
    and environmental differentiation in the carnivorous plant *Serracenia alata*. Syst. Biol.
    61:763-777.

Zerbino D. R., Birney E. 2008. Velvet: Algorithms for de novo short read assembly using de
    Bruijn graphs. Genome Res. 18:821-829.

Zhang, C., Zhang, D. X., Zhu, T.,  Yang, Z. 2011. Evaluation of a Bayesian coalescent method of
    species delimitation. Syst. Biol. 60: 747-761.

Table 1. UCE datasets used for computation of summary statistics and subsequent analyses.

| Dataset Name | Number | Number of Loci | Number of Individuals in Alignments | Description | Analyses |
|---|---|---|---|---|---|
| Single Species Full | 5 (one per species) | Variable (776-1,516) | 8 (4 for *M. marginatus*) | All loci within a given species that have data for at least one individual in each area of endemism | Summary statistics, G-PhoCS |
| Single Species Reduced | 5 (one per species) | 166 | 8 (4 for *M. marginatus*) | All loci within a given species that fulfill the criterion for the full dataset for all five study species | Summary statistics, *BEAST, G-PhoCS, BP&P |
| All Species Full | 1 | 2,219 | 36 | All loci that are present in at least two individuals across species | Summary statistics |
| All Species Reduced | 1 | 169 | 36 | All loci that that have data for at least one individual in each area of endemism for all species | Summary statistics |

Table 2. Mean values (±SD) of population genetic summary statistics Watterson's θ, nucleotide diversity (π), and Tajima's *D* from both UCE single species datasets and alignments of the mitochondrial gene Cytochrome *b*.

| Scientific name | Full UCE | | | Reduced UCE | | | Cytochrome *b* | | |
|---|---|---|---|---|---|---|---|---|---|
| | $\theta_W$ | π | Tajima's *D* | $\theta_W$ | π | Tajima's *D* | $\theta_W$ | π | Tajima's *D* |
| *Cymbilaimus lineatus* | 0.0021 (0.0061) | 0.0019 (0.0055) | -0.4886 (0.7282) | 0.0024 (0.0070) | 0.0021 (0.0064) | -0.5248 (0.6846) | 0.0119[b] | 0.0133[b] | 0.6371[b] |
| *Xenops minutus* | 0.0021 (0.0045) | 0.0019 (0.0040) | -0.2241 (0.8608) | 0.0018 (0.0028) | 0.0017 (0.0031) | -0.3089 (0.8809) | 0.0355[b] | 0.0450[b] | 1.44979[b] |
| *Schiffornis turdina* | 0.0029 (0.0042) | 0.0025 (0.0038) | -0.6353 (0.6476) | 0.0022 (0.0028) | 0.0020 (0.0030) | -0.6044 (0.6351) | 0.0341[b] | 0.0351[b] | 0.1669[b] |
| *Querula purpurata* | 0.0014 (0.0045) | 0.0013 (0.0048) | -0.5382 (0.7060) | 0.0011 (0.0015) | 0.0009 (0.0015) | -0.6624 (0.7257)[a] | 0.0041 | 0.0045 | 0.5555[b] |
| *Microcerculus marginatus* | 0.0015 (0.0021) | 0.0015 (0.0021) | -0.3036 (0.6188) | 0.0012 (0.0019)[a] | 0.0011 (0.0018)[a] | -0.3480 (0.5662) | 0.0291[b] | 0.0267[b] | -0.8690[b] |

[a] = significantly different from equivalent summary statistic calculated from non-orthologous UCEs at p < 0.05 level (Mann-Whitney *U* test)

[b] = value falls outside 95% confidence limits of equivalent summary statistic calculated from all UCEs

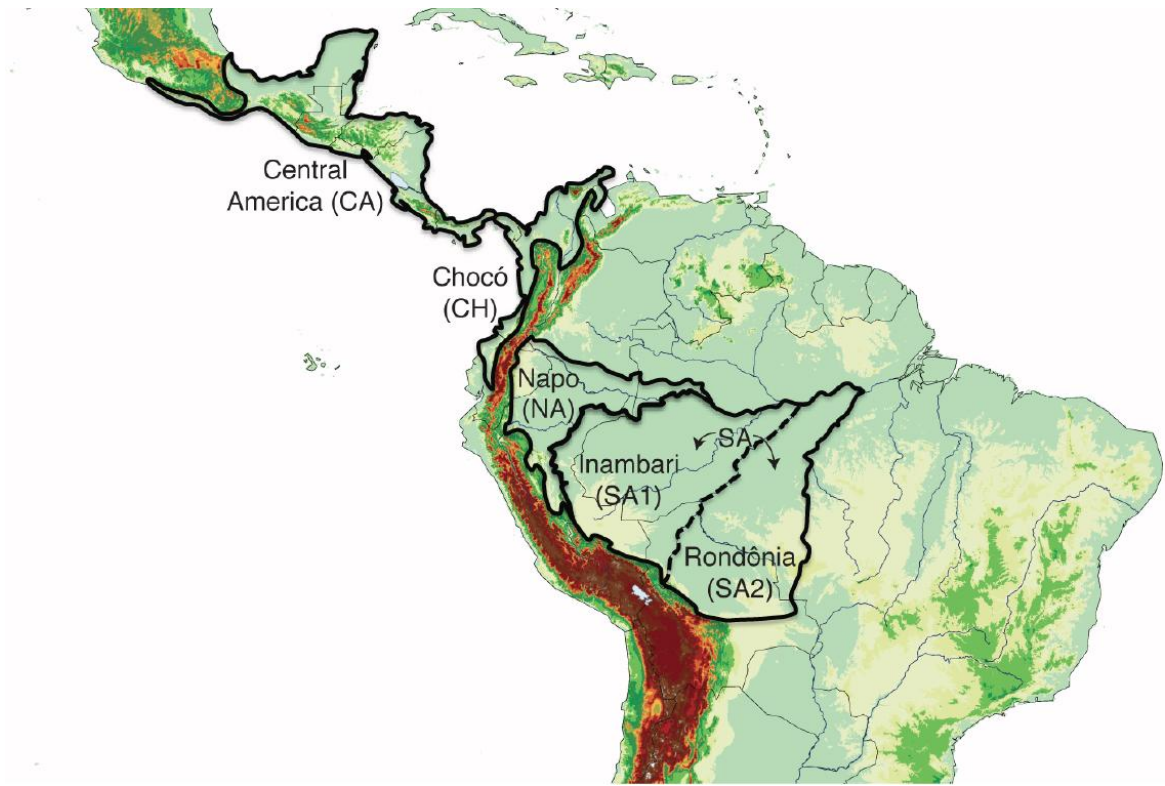

Figure 1. Map of the areas of endemism for lowland Neotropical birds that we used to define populations for this study. For *C. lineatus*, *X. minutus*, and *Q. purpurata*, Inambari (SA1) and Rondônia (SA2) were collapsed as SA.

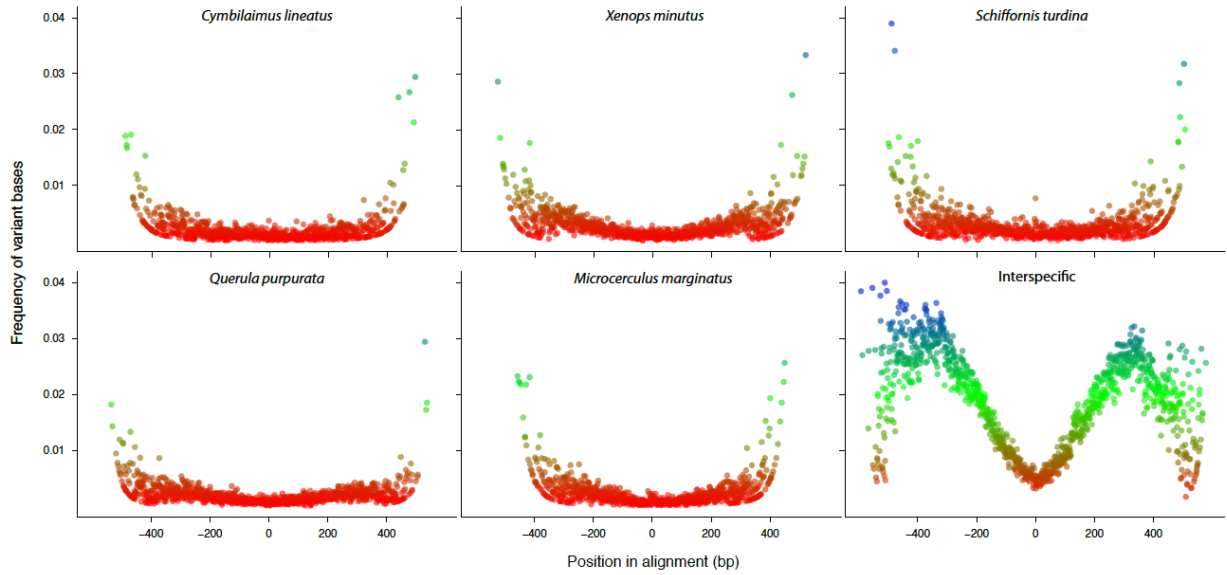

Figure 2. Variability increases in flanking regions of ultraconserved regions within each of the five study species (single species full datasets) and in alignments of all individuals across all five study taxa (all species full dataset). We removed data points with no variability and outliers for clarity of presentation. The apparent decrease in variability toward the edges in the interspecific alignments likely results from differences in alignment lengths and missing/reduced data near the ends of alignments, although reduced sequence coverage in these outlying areas may also reduce our ability to accurately call variants.

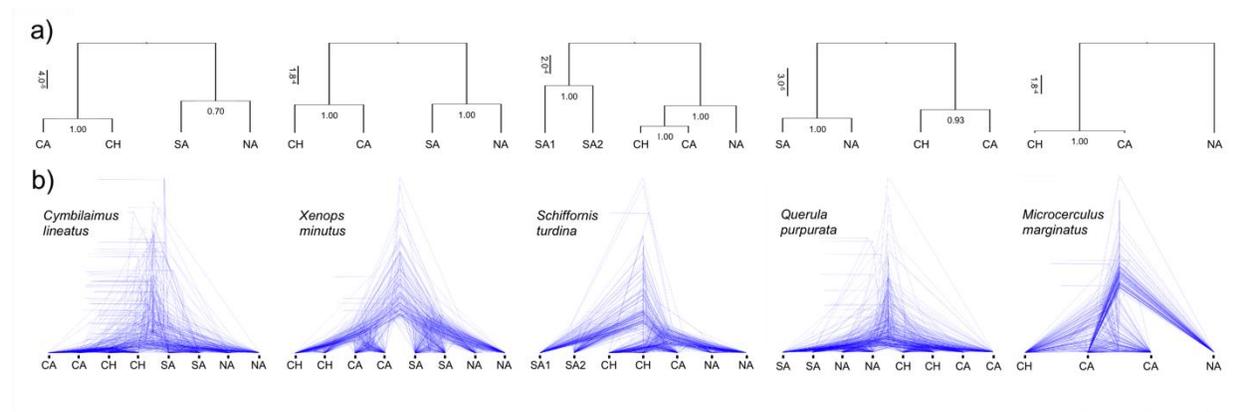

Figure 3. (a) Maximum Clade Credibility species trees with posterior probability values for each node and (b) cloudograms showing the Maximum Clade Credibility tree for each UCE from *BEAST analyses of the single species reduced datasets.

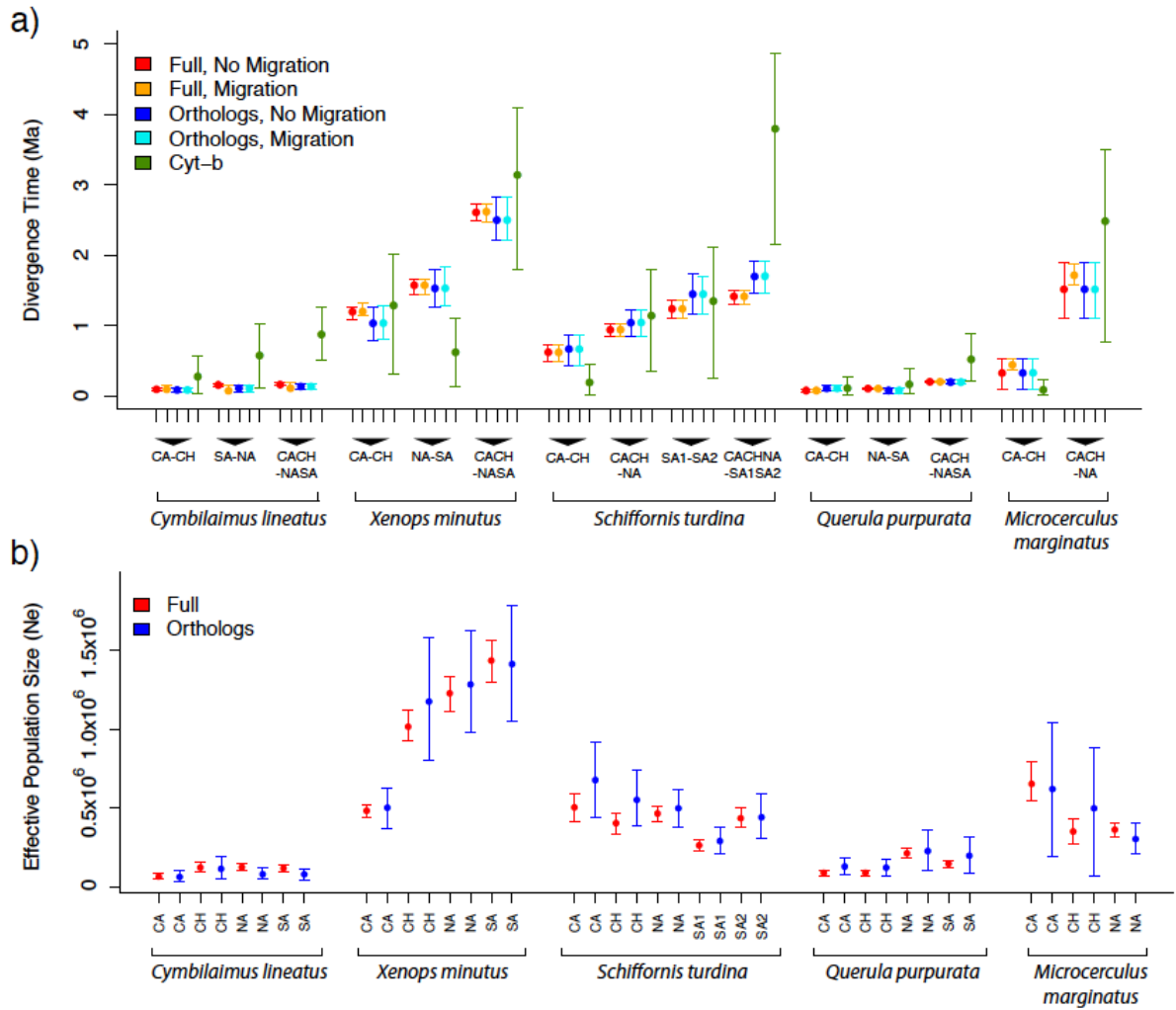

Figure 4. Results from G-PhoCS demographic modeling including (a) divergence times for each population divergence event in each study species based on output from G-PhoCS. Results of all four models are presented: ($m_3$) single species full, no migration; ($m_4$) single species full, migration; ($m_1$) single species reduced, no migration; and ($m_2$) single species reduced, migration. Divergence events are labeled by listing the populations included in both of the resulting clades. (b) Effective population sizes ($N_e$) for all extant populations in each study species. Results of two models are presented: ($m_4$) single species full, migration and ($m_2$) single species reduced, migration.

**SUPPLEMENTARY METHODS**

*Sanger sequencing and mtDNA*

In order to compare our UCE results to patterns inferred from a more traditional marker used in phylogeographic studies, we constructed a complimentary mtDNA data set using the gene cytochrome *b* for the five species. We sampled 36 individuals from five species - *Cymbilaimus lineatus, Xenops minutus, Schiffornis turdina, Querula purpurata,* and *Microcerculus marginatus.* We extracted DNA from ~ 25 mg of pectoral muscle tissue using the DNeasy Tissue Kit (Qiagen, Valencia, CA). We performed polymerase chain reaction (PCR) amplifications (25 mL) of cyt *b* using primers L14764 (Sorenson et al., 1999) and H4A (Harshman, 1996). Our reaction consisted of consisted of 2.5 mL template DNA (~50 ng), 0.3 mL of each primer (10 mM), 0.5 mL dNTPs (10 mM), 2.5 mL 10X with MgCl2 reaction buffer, 0.1 Taq DNA polymerase (5 U/mL AmpliTaq; ABI, Foster City, CA), and 18.7 mL sterile dH2O. PCR temperature profiles consisted of an initial denaturation of 2 min at 94"C, followed by 35 cycles of 30 s at 94°C, 30 s at 45°–48°C, and 2 min at 72°C, with a final extension of 5 min at 72°C. We purified double-stranded PCR products using 20% polyethylene glycol and then we performed a cycle-sequence reaction using 1.75 mL 5X sequencing buffer (ABI), 1 mL sequencing primer (10 mM), 2.25 mL template, 0.35 mL Big Dye Terminator Cycle-Sequencing Kit (ver.3.1; ABI), and 1.65 mL sterile dH2O, for a total volume of 7 mL. We cleaned reactions using Sephadex (G-50 fine) columns and analyzed on an ABI 3100 Genetic Analyzer. Finally, we manually assembled contigs for each individual using Sequencher (ver. 4.9; GeneCodes, Ann Arbor, MI), and we verified base calls by eye and that there were no stop codon in the coding region.

**REFERENCES**

Harshman J., 1996. Phylogeny, evolutionary rates, and ducks. Ph.D. dissertation, University of Chicago, Chicago.

Sorenson M.D., Ast J.C., Dimcheff D.E., Yuri T., Mindell D.P. 1999. Primers for a PCR-based approach to mitochondrial genome sequencing in birds and other vertebrates. Mol. Phylogenet. Evol.12: 105–114.

Supplementary Table 1. Samples used in test of UCEs for phylogeography. The number of trimmed reads produced per sample and the number of alignments used in the full datasets are also included.

| Taxon[a] | Museum tissue no. | Geographic Region[b] | Latitude | Longitude | Number of Trimmed Reads | Representation in Alignments |
|---|---|---|---|---|---|---|
| *Cymbilaimus lineatus* (Fasciated Antshrike) | | | | | | |
| *C. l. fasciatus* | UNLV GMS1038 | CA | 8.5107 | -81.1165 | 1148600 | 560 |
| *C. l. fasciatus* | LSUMZ B28582 | CA | 9.2083 | -79.9955 | 3299807 | 766 |
| *C. l. fasciatus* | LSUMZ B2205 | CH | 7.7560 | -77.6840 | 4233705 | 773 |
| *C. l. fasciatus* | LSUMZ B2252 | CH | 7.7560 | -77.6840 | 2136887 | 740 |
| *C. l. intermedius* | LSUMZ B4157 | NA | -2.8200 | -73.2738 | 2014330 | 732 |
| *C. l. intermedius* | LSUMZ B6890 | NA | -3.3137 | -72.5200 | 1909182 | 724 |
| *C. l. intermedius* | MPEG PUC209 | SA | -4.8500 | -65.0667 | 1351532 | 668 |
| *C. l. intermedius* | LSUMZ B1129 | SA | -15.5000 | -67.3000 | 1202104 | 745 |
| *Xenops minutus* (Plain Xenops) | | | | | | |
| *X. m. mexicanus* | KUNHM 2044 | CA | 18.5928 | -90.2561 | 11876932 | 1338 |
| *X. m. mexicanus* | LSUMZ B60935 | CA | 14.8728 | -87.9050 | 11803798 | 1333 |
| *X. m. littoralis* | LSUMZ B2209 | CH | 7.7560 | -77.6840 | 14180494 | 1360 |
| *X. m. littoralis* | LSUMZ B11948 | CH | 0.8667 | -78.5500 | 15060302 | 1356 |
| *X. m. obsoletus* | LSUMZ B4244 | NA | -2.8200 | -73.2738 | 12010139 | 1335 |
| *X. m. obsoletus* | LSUMZ B6862 | NA | -3.4167 | -72.5833 | 8801265 | 1330 |
| *X. m. obsoletus* | FMNH 433364 | SA | -13.0167 | -71.4833 | 9616279 | 1363 |
| *X. m. obsoletus* | LSUMZ B9026 | SA | -11.4703 | -68.7786 | 10375326 | 1334 |
| *Schiffornis turdina* (Brown-winged Schiffornis) | | | | | | |
| *S. t. dumicola* | UNLV GMS1112 | CA | 8.6330 | -80.1058 | 2142820 | 740 |
| *S. t. panamensis* | UNLV JMD85 | CA | 9.2397 | -79.4127 | 5872626 | 846 |
| *S. t. rosenbergi* | LSUMZ B11889 | CH | 0.8667 | -78.5500 | 1892422 | 819 |
| *S. t. rosenbergi* | LSUMZ B11820 | CH | 0.8667 | -78.5500 | 6633643 | 838 |
| *S. t. amazona/aenea*[c] | LSUMZ B27903 | NA | -7.0833 | -75.6500 | 5429052 | 829 |
| *S. t. amazona/aenea*[c] | LSUMZ B27769 | NA | -7.0833 | -75.6500 | 6515331 | 839 |
| *S. t. steinbachi* | LSUMZ B22835 | SA1 | -15.1881 | -68.2550 | 5161296 | 851 |
| *S. t. subsp. nov.* | LSUMZ B14888 | SA2 | -14.5000 | -60.6500 | 3751586 | 851 |
| *Querula purpurata* (Purple-throated Fruitcrow) | | | | | | |
| *Q. p.* | LSUMZ B1955 | CA | 8.9501 | -82.1200 | 6253720 | 1482 |
| *Q. p.* | LSUMZ B28529 | CA | 9.2083 | -79.9955 | 2775149 | 864 |
| *Q. p.* | LSUMZ B2217 | CH | 7.7560 | -77.6840 | 2335868 | 866 |
| *Q. p.* | LSUMZ B2219 | CH | 7.7560 | -77.6840 | 11955357 | 1490 |
| *Q. p.* | LSUMZ B2823 | NA | -3.1793 | -72.9035 | 8959400 | 1468 |
| *Q. p.* | LSUMZ B103546 | NA | -5.0833 | -74.5833 | 11083592 | 1491 |
| *Q. p.* | KUNHM 1419 | SA | -12.5500 | -69.0500 | 10640666 | 1499 |
| *Q. p.* | LSUMZ B9648 | SA | -11.4703 | -68.7786 | 3616142 | 1408 |
| *Microcerculus marginatus* (Scaly-breasted Wren) | | | | | | |
| *M. m. luscinia* | LSUMZ 28551 | CA | 9.2083 | -79.9955 | 8802513 | 1034 |
| *M. m. luscinia* | LSUMZ 28555 | CA | 9.2083 | -79.9955 | 10351778 | 1065 |
| *M. m. luscinia* | LSUMZ 2290 | CH | 7.7560 | -77.6840 | 12761364 | 1077 |
| *M. m. marginatus* | LSUMZ 2662 | NA | -3.1793 | -72.9035 | 11349173 | 1077 |

[a] taxonomy based on Dickinson (2003)

[b] CA = Central America, CH = Chocó, NA = Napo, SA = Inambari/Rondônia

[c] specimen from this locality appears intermediate between *aenea* and *amazona*

Supplementary Table 2. Custom blocking DNA sequences used in target enrichment of UCEs.

| Name | Sequence |
| --- | --- |
| revcomp-univ-adapter | 5' - AGATCGGAAGAGCGTCGTGTAGGGAAAGAGTGTAGATCTCGGTGGTCGCCGTATCATT - 3' |
| revcomp-index-adapter-sequence1 | 5' - CAAGCAGAAGACGGCATACGAGAT  - 3' |
| revcomp-index-adapter-sequence2 | 5' - GTGACTGGAGTTCAGACGTGTGCTCTTCCGATCT - 3' |

Supplementary Table 3. Summary of assembly and alignment statistics from VELVET and MAFFT.

| Common Name | Trimmed Reads | Reads in Contigs | % Trimmed Reads in Contigs | Total Contigs | Average Coverage | Average Contig Length | Contigs Mapping to UCE | MAFFT Alignments | Final Filtered Alignments |
|---|---|---|---|---|---|---|---|---|---|
| Fasciated Antshrike | 17296147 | 4573979 | 26.4% | 2899 | 386.56 | 323.03 | 951 | 812 | 776 |
| Plain Xenops | 93724535 | 6323254 | 6.7% | 6481 | 284.65 | 297.70 | 1512 | 1426 | 1368 |
| Thrush-like Schiffornis | 37398776 | 8771936 | 23.5% | 6698 | 361.71 | 257.61 | 1066 | 895 | 851 |
| Purple-throated Fruitcrow | 57619894 | 6897426 | 12.0% | 10006 | 181.85 | 304.69 | 1796 | 1580 | 1516 |
| Southern Nightingale-Wren | 43264828 | 2920120 | 6.7% | 2545 | 334.64 | 371.26 | 1190 | 1108 | 1077 |
| Average | 49860836 | 5897343 | 15.1% | 5726 | 309.88 | 310.86 | 1303 | 1164 | 1118 |
| Total | 249304180 | 29486715 | - | 28629 | - | - | 6515 | 5821 | 5588 |

Supplementary Table 4. Proportion of polymorphic UCE loci and variable sites per variable locus in UCE single species datasets.

| Scientific name | Proportion UCE Loci Polymorphic | | Average Variable Sites per Variable UCE Locus | |
|---|---|---|---|---|
| | *Full* | *Reduced* | *Full* | *Reduced* |
| *Cymbilaimus lineatus* | 0.5271 | 0.5060 | 2.3692 | 2.3571 |
| *Xenops minutus* | 0.7300 | 0.7229 | 2.8982 | 2.8750 |
| *Schiffornis turdina* | 0.7709 | 0.7651 | 3.2304 | 3.1654 |
| *Querula purpurata* | 0.5797 | 0.5904 | 2.0906 | 2.3776 |
| *Microcerculus marginatus* | 0.5993 | 0.5301 | 1.9970 | 1.7159 |

Supplementary Table 5. Relative substitution rates for UCEs. Shown are the net genetic distance (π) within a species, which are the mean distances across the full and reduced UCE datasets and cytochrome *b*. Calculations are based on the single species full and single species reduced datasets. Relative substitution ratios (UCE π/ Cytb π) and relative substitution rates for UCEs (UCE π/ CytB π x 0.0105 (sub/site/million years)) are provided. The Cytb substitution rate is based on a published molecular clock for birds. Relative substitution rates for UCE are shown for the reduced and full datasets.

| Taxon | Full UCE π | Reduced UCE π | Cytb π | Reduced UCE Relative Sub Ratio | Cytb Sub Rate (sub/site/ million years) | Relative Reduced UCE Sub Rate (sub/site/ million years) | Full UCE Relative Sub Ratio | Relative Full UCE Sub Rate (sub/site/ million years) |
|---|---|---|---|---|---|---|---|---|
| *Cymbilaimus lineatus* | 0.0021 | 0.0019 | 0.0133 | 0.1579 | 0.0105 | 0.0017 | 0.1429 | 0.0015 |
| *Xenops minutus* | 0.0017 | 0.0019 | 0.0450 | 0.0378 | 0.0105 | 0.0004 | 0.0422 | 0.0004 |
| *Schiffornis turdina* | 0.0020 | 0.0025 | 0.0351 | 0.0570 | 0.0105 | 0.0006 | 0.0712 | 0.0007 |
| *Querula purpurata* | 0.0009 | 0.0013 | 0.0045 | 0.2000 | 0.0105 | 0.0021 | 0.2889 | 0.0030 |
| *Microcerculus marginatus* | 0.0011 | 0.0015 | 0.0267 | 0.0412 | 0.0105 | 0.0004 | 0.0562 | 0.0006 |

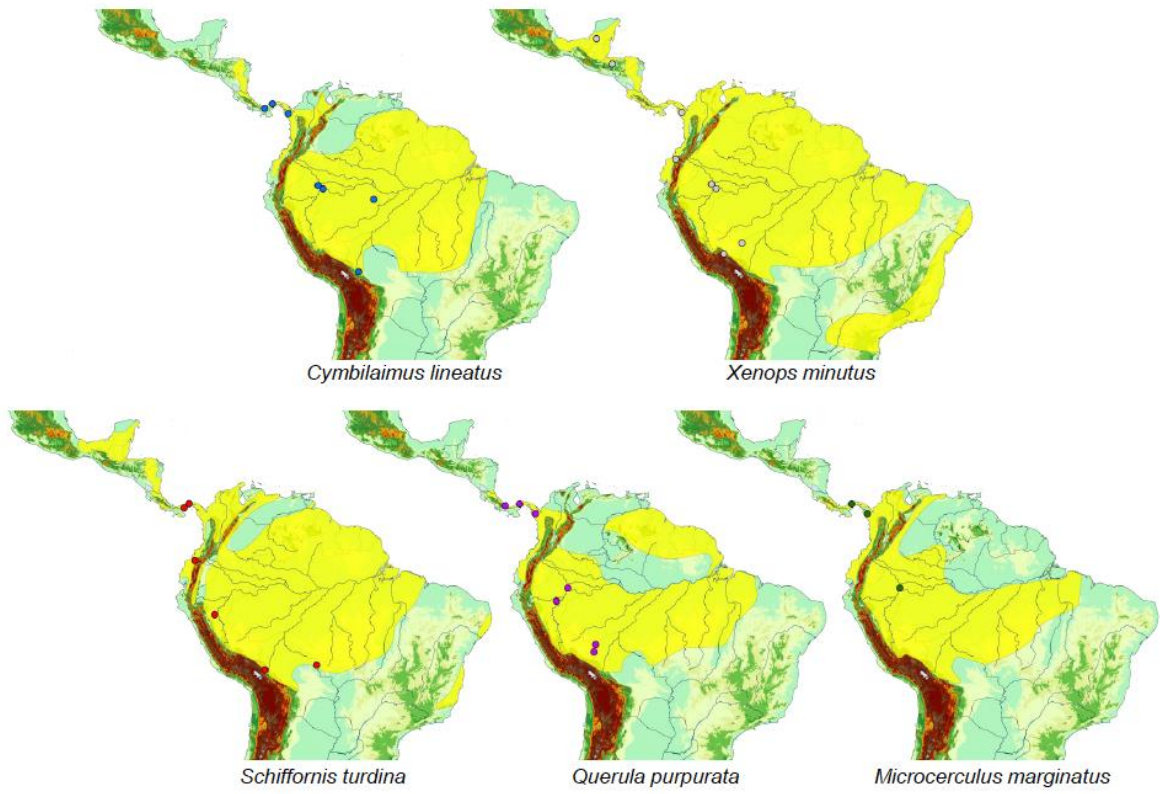

*Cymbilaimus lineatus*

*Xenops minutus*

*Schiffornis turdina*

*Querula purpurata*

*Microcerculus marginatus*

Supplementary Figure 1. Range maps showing the geographic distribution of each study species (in yellow). Sampling localities are marked with colored circles.

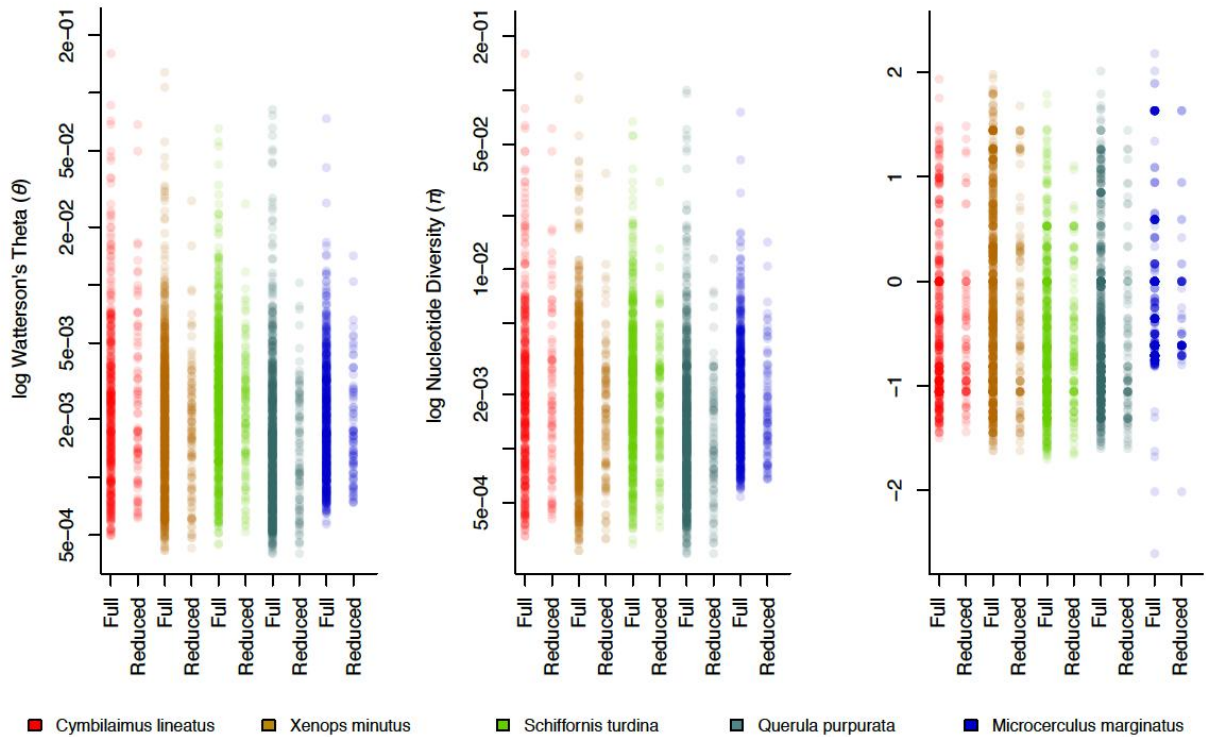

Supplementary Figure 2. Population genetic summary statistics calculated from single species full and single species reduced datasets. We removed non-polymorphic loci, for presentation on a log axis for Watterson's θ and nucleotide diversity and because Tajima's *D* can only be calculated from polymorphic loci.

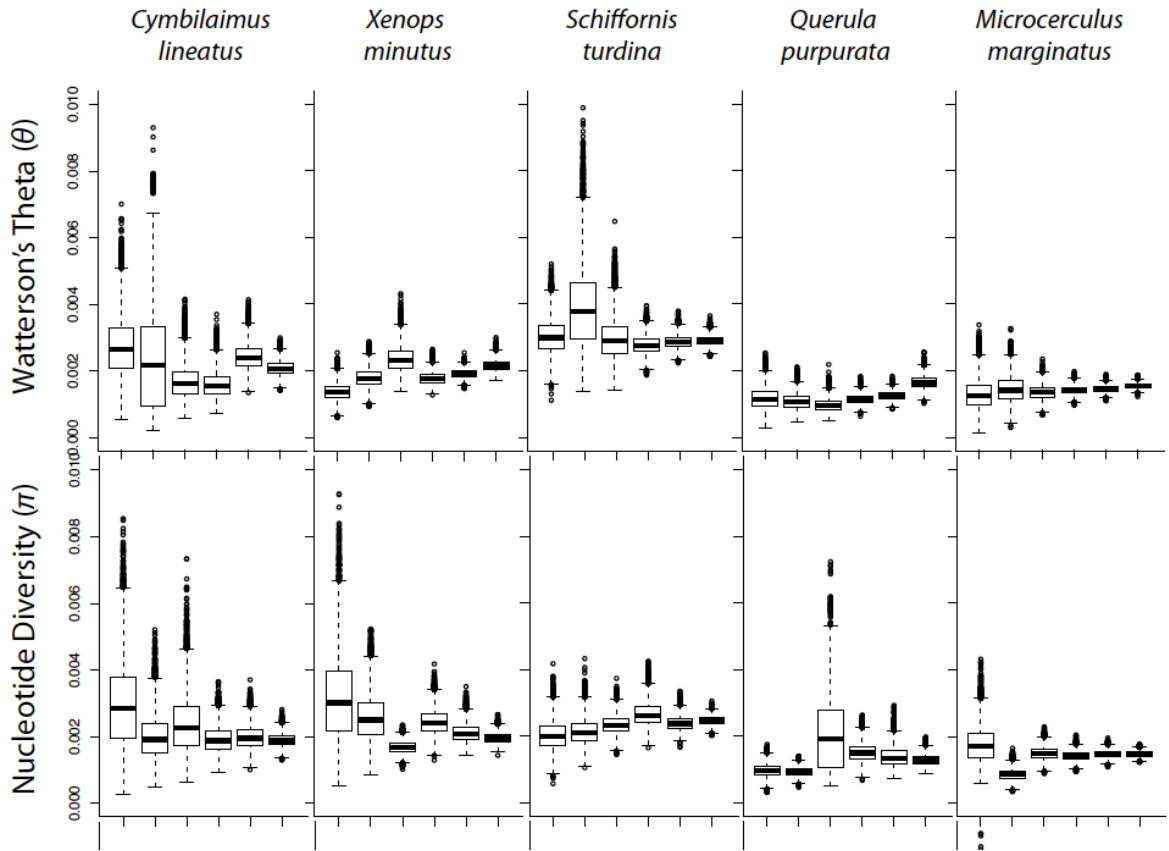

Supplementary Figure 3. Boxplots of summary statistics based on 10,000 bootstrap replicates from subsamples of the single species full datasets containing different numbers of loci, revealing less variance in estimates derived from larger samples of loci.

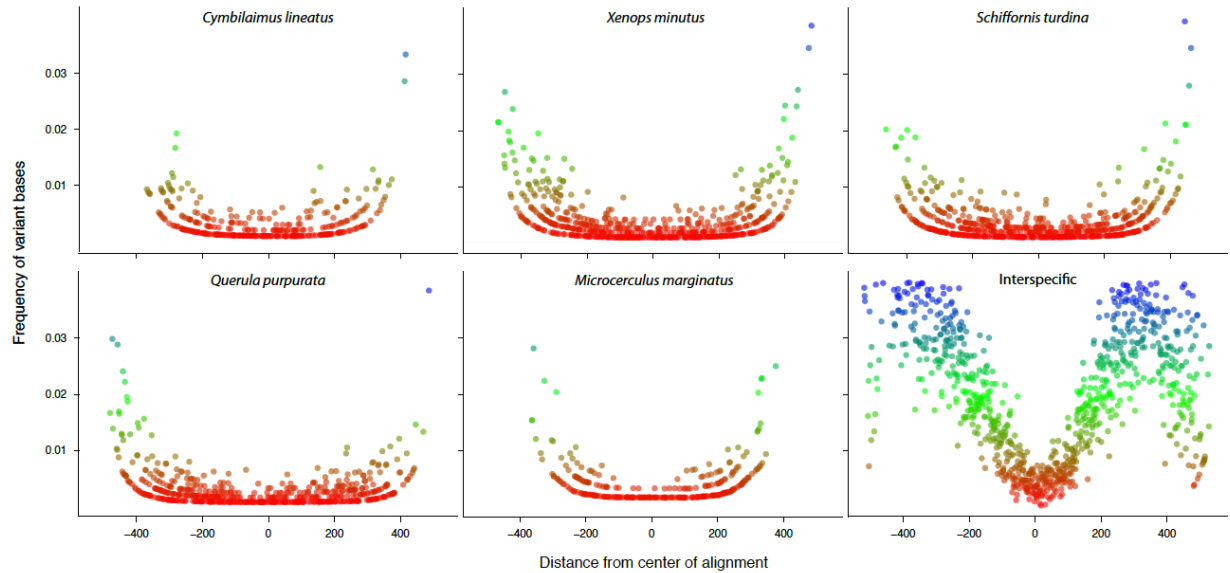

Supplementary Figure 4. Variability increases in flanking regions of ultraconserved regions in all five study species (single species reduced datasets) and in alignments of all individuals from all five study taxa (all species reduced dataset). We removed data points with no variability and outliers for clarity of presentation. The apparent decrease in variability toward the edges in the interspecific alignments likely results from differences in alignment lengths and missing/reduced data near the ends of alignments, although reduced sequence coverage in these outlying area may also reduce our ability to accurately call variants.

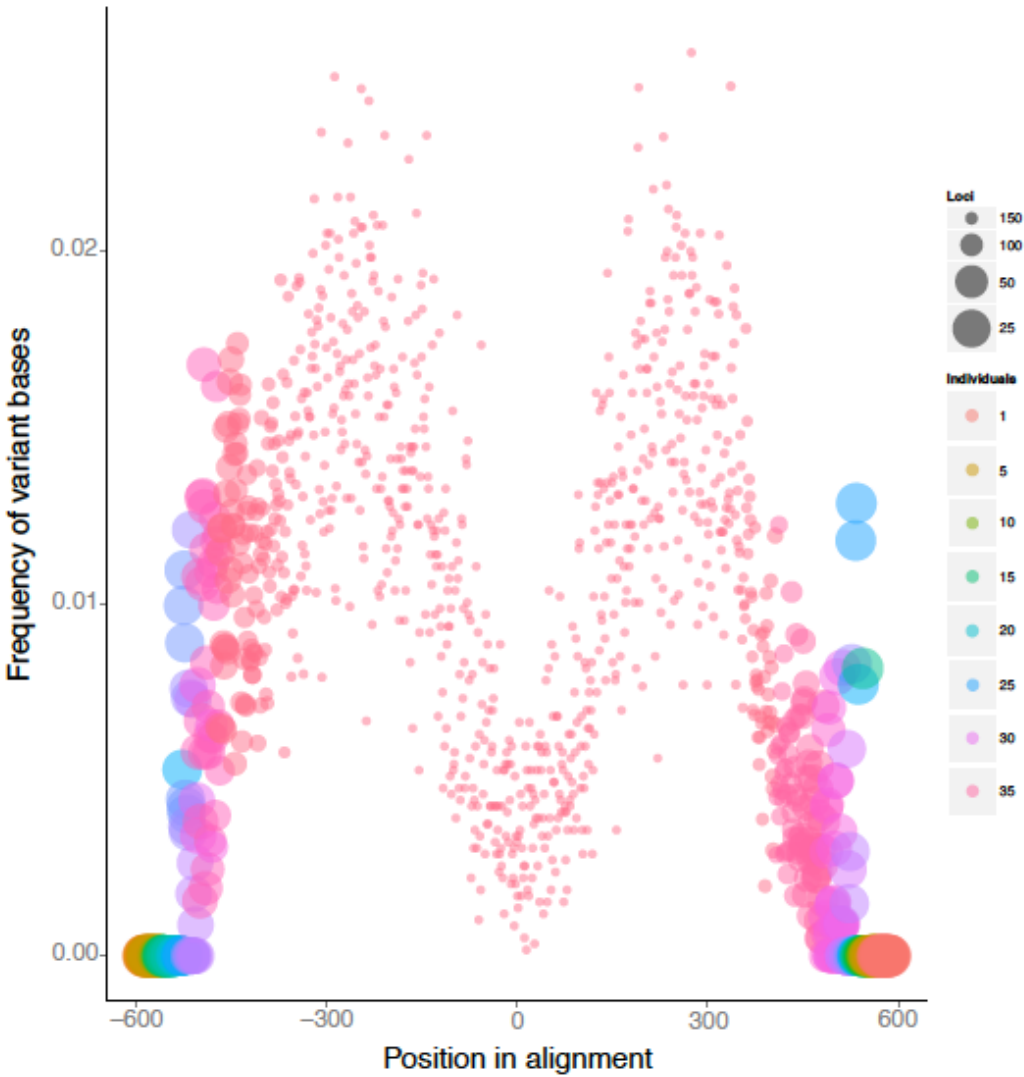

**Supplementary Figure 5.** The frequency of variant bases by alignment position where data points are scaled by size to indicate the number of loci having data at each position and data points vary by color to indicate the number of individuals having data at each position. Variability initially increases moving from the core of UCE regions into their flanking sequence. Variability decreases beyond approximately 375 bp from center because there are fewer alignments of this length and because alignments having data in outer flanks are comprised of fewer individuals. The decrease in variability in outer flanks may also result from reduced read depth across these positions.

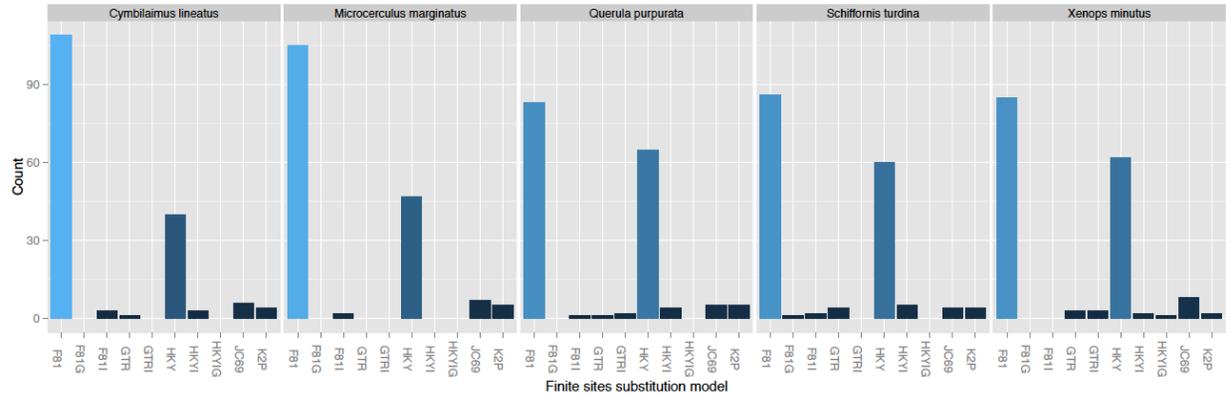

Supplementary Figure 6: The best-fit finite sites substitution model for all UCEs in each species based on CloudForest analyses.

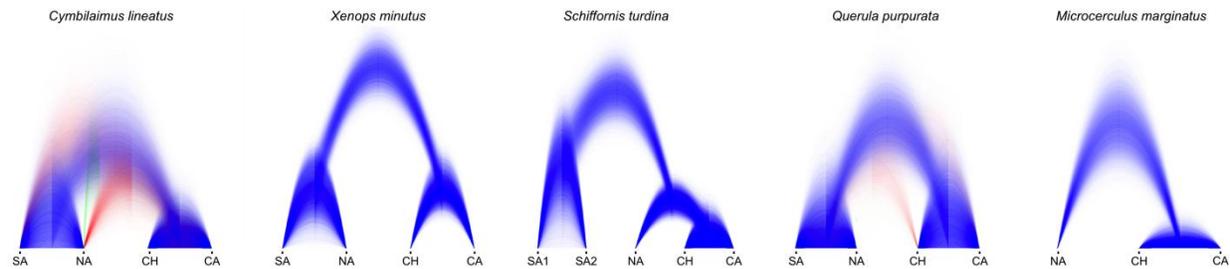

Supplementary Figure 7. Cloudograms illustrating the posterior distribution of species trees for each study taxon, based on *BEAST analyses of single species reduced datasets. Trees concordant with the MCC trees are colored blue whereas trees discordant with the MCC trees are colored red or green.

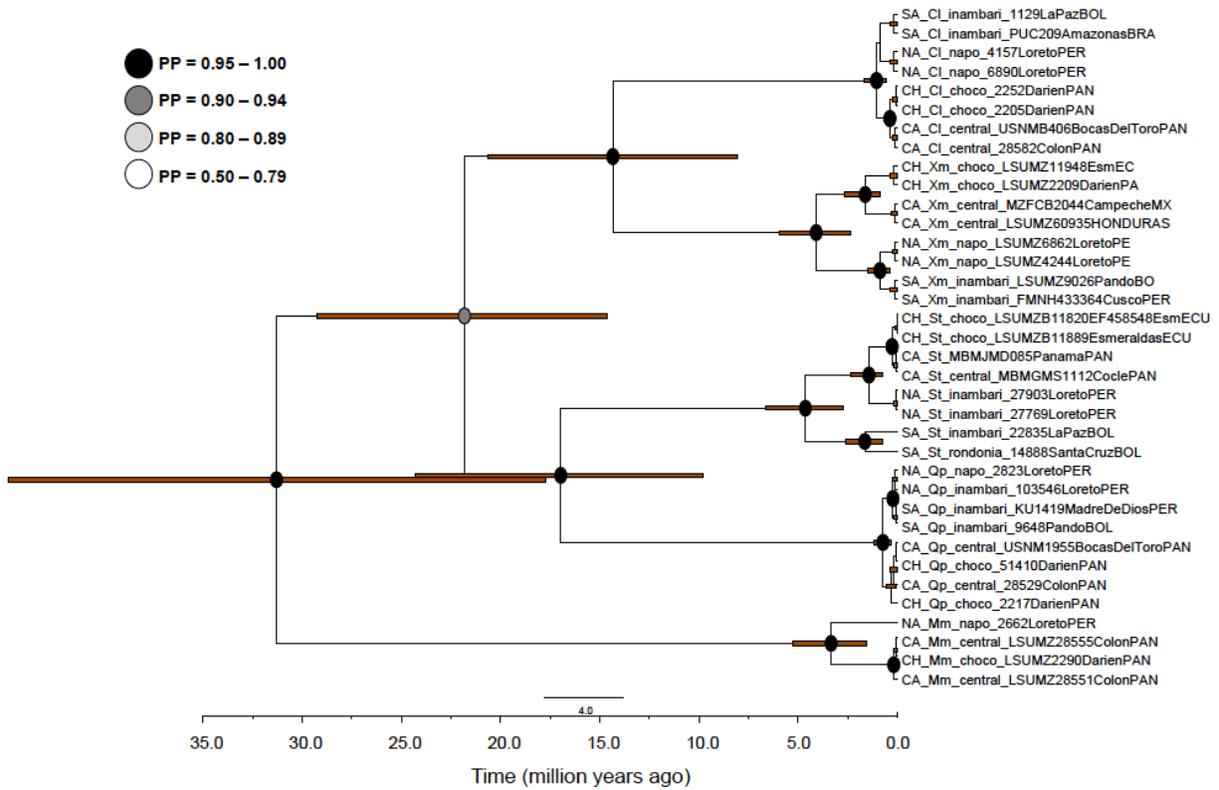

Supplementary Figure 8. A Maximum Clade Credibility (MCC) and time-calibrated gene tree constructed from the mitochondrial cytochrome *b* gene using BEAST. Nodes are color-coded according to their posterior probability (PP) and confidence intervals are provided for node ages.

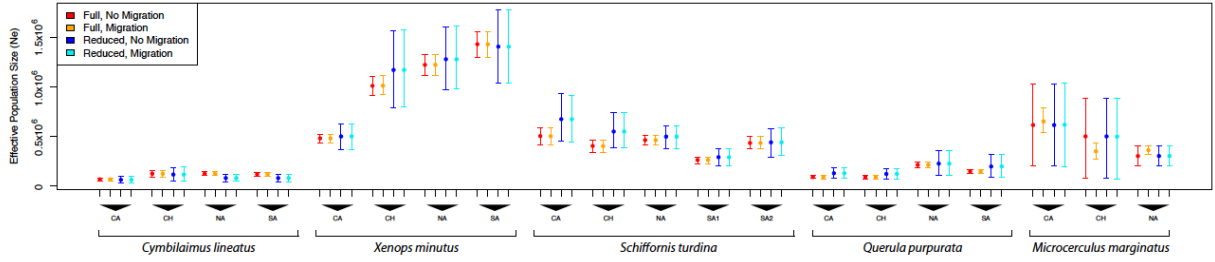

Supplementary Figure 9. Plot of effective population sizes ($N_e$) for all terminal populations in each study species. Results of all four models are presented: ($m_3$) single species full, no migration; ($m_4$) single species full, migration; ($m_1$) single species reduced, no migration; and ($m_2$) single species reduced, migration.

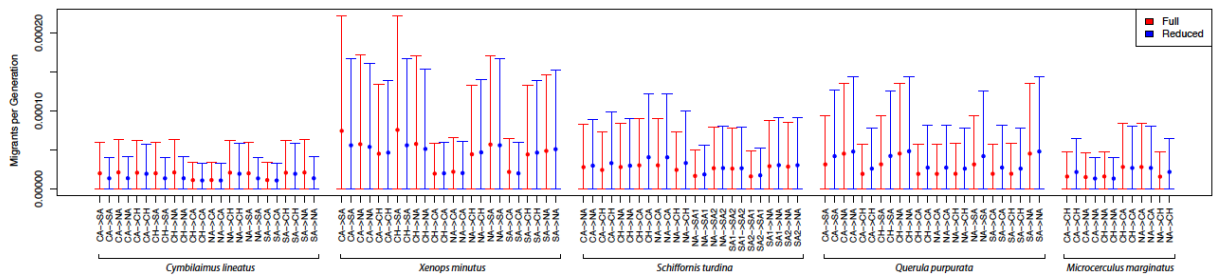

Supplementary Figure 10. Plot of migration rates (migrants per generation) between terminal populations in each study species. Results of the two models incorporating migration are presented: ($m_4$) single species full, migration and ($m_2$) single species reduced, migration. Arrows in labels indicate the direction of migration.

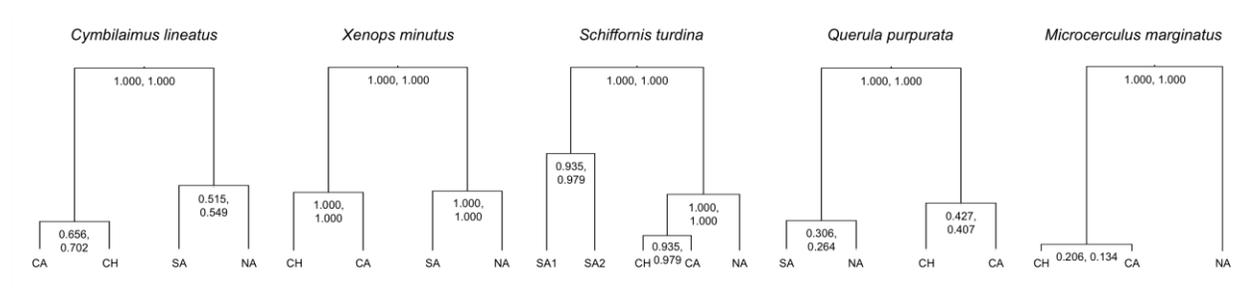

Supplementary Figure 11. Speciation probabilities for each terminal lineage for each study species based on results from BP&P and single species reduced datasets. Probabilities are provided using two prior distributions for the θ (theta) and τ (tau) gamma priors: 1) θ = (1, 30) and τ = (1, 30) and 2) θ = (1, 300) and τ = (1, 300).